\documentclass[prd,twocolumn,floatfix,amsmath,nofootinbib,amssymb,floatfix]{revtex4}
\usepackage{graphicx,color,dcolumn,booktabs,bm}
\usepackage{longtable,lscape}
\usepackage{txfonts}
\usepackage{overpic}
\usepackage{amssymb}
\usepackage{indentfirst}
\usepackage{feynmf}   %{feynmp}
\usepackage{slashed}  %for Feynman symbols
\usepackage{cases}
\usepackage{color}
\usepackage{multirow}
\usepackage{slashed}
\usepackage{epstopdf}
\usepackage{longtable}
\usepackage{ulem}
\usepackage{graphicx,color,dcolumn,booktabs,bm}
\usepackage[colorlinks,
            citecolor=blue,
            anchorcolor=red,
            menucolor=red,
            linkcolor=red,
            filecolor=red,
            runcolor=red,
            urlcolor=blue,
            frenchlinks=red]{hyperref}

\graphicspath{{Figures/}} %
\allowdisplaybreaks

\begin{document}

\title{Light unflavored vector meson spectroscopy around the mass range of $2.4\sim 3$ GeV and possible experimental evidence}

\author{Li-Ming Wang$^{1,3}$}\email{lmwang@ysu.edu.cn}
\author{Si-Qiang Luo$^{2,3,4}$}\email{luosq15@lzu.edu.cn}
\author{Xiang Liu$^{2,3,4,5}$\footnote{Corresponding author}}\email{xiangliu@lzu.edu.cn}
 \affiliation{
 $^1$Key Laboratory for Microstructural Material Physics of Hebei Province, School of Science, Yanshan University, Qinhuangdao 066004, China\\
$^2$School of Physical Science and Technology, Lanzhou University, Lanzhou 730000, China\\
$^3$Lanzhou Center for Theoretical Physics, Key Laboratory of Theoretical Physics of Gansu Province and Frontiers Science Center for Rare Isotopes, Lanzhou University, Lanzhou 730000, China\\
$^4$Research Center for Hadron and CSR Physics, Lanzhou University and Institute of Modern Physics of CAS, Lanzhou 730000, China\\
$^5$Joint Research Center for Physics, Lanzhou University and Qinghai Normal University, Xining 810000, China}
\begin{abstract}
In this work, we predict the spectroscopy behavior of these light unflavored vector mesons with masses at the range of $2.4\sim 3$ GeV, which are still missing in the experiment. By presenting their mass spectrum and studying their two-body Okubo-Zweig-lizuka allowed decay widths, we discuss the possible experimental evidences of these discussed states combining with the present experimental data. Especially, we strongly suggest that our experimental colleagues carry out the exploration of these higher states via the $e^+e^-$ annihilation into light mesons. It is obvious that BESIII and Belle II will be potential experiments to achieve this target.

\end{abstract}
%\pacs{14.40.Be, 12.38.Lg, 13.25.Jx}
\date{\today}
\maketitle

%%%%%%%%%%%%%%%%%%%%%%%%%%%%%%%%%%%%%%%%%%%%%%%%%%%%%%
\section{Introduction}\label{sec1}
%%%%%%%%%%%%%%%%%%%%%%%%%%%%%%%%%%%%%%%%%%%%%%%%%%%%%%

As a main body of the hadron family, light hadron spectroscopy has special status in the whole hadron family, which has inspired the classification of hadron based on SU(3) symmetry in 1964 \cite{Gell-Mann:1964ewy,Zweig:1964ruk}, and has aroused the interest from both theorists and experimentalists in exploring exotic states like glueball, hybrid, and multiquark states \cite{Teper:1998kw,Meyer:2015eta,Chen:2016qju,Liu:2019zoy}. With the running of BESIII and Belle II, it is a good chance to launch a profound research on light flavor mesons, which can deepen our understanding of how quarks and gluons interact with each other to form hadrons.

Focusing on the study of light unflavored vector meson spectroscopy, we find big progress being made by experiments in the past years \cite{Ablikim:2005um,Ablikim:2010au,Ablikim:2016itz}. A typical example is the observation of the $Y(2175)$ \cite{Ablikim:2007ab} and the following measurement \cite{Ablikim:2014pfc,Lees:2011zi,Aubert:2007ym,Shen:2009zze,Aubert:2007ur,Aubert:2006bu} around the $Y(2175)$, which have stimulated extensive discussions of decoding the property of the $Y(2175)$ \cite{Wang:2021gle,Wang:2006ri,Ding:2007pc,MartinezTorres:2008gy,Chen:2008ej,Klempt:2007cp}. Among these possible explanations to the $Y(2175)$, categorizing the $Y(2175)$ into the $\phi$ meson family is a popular one \cite{Wang:2012wa,Ding:2007pc,Barnes:2002mu}, which should still be tested in an experiment. In fact, the light unflavored vector mesons also include these states in the $\rho$ and $\omega$ meson families. With the accumulation of data with unprecedented statistical accuracy, a series of $\rho$ and $\omega$ states around 2 GeV were observed in experiments \cite{Antonelli:1996xn,Clegg:1989mp,Hasan:1994he,Bugg:2004xu,Atkinson:1985yx,Chung:2002pu,Anisovich:2011sva,Bugg:2004rj,Atkinson:1988xa}. By the joint effort from both theorists and experimentalists, our knowledge of light unflavored vector meson spectroscopy below 2.2 GeV has been promoted, which is the situation of the study of unflavored vector meson spectroscopy. But, it is not the whole aspect of light unflavored vector meson spectroscopy.

The experiments have released the data of $e^+e^-$ annihilation into light mesons \cite{BaBar:2006vzy,BESIII:2019gjz,BaBar:2011btv,Belle:2008kuo}, where the collision energy reaches up to $\sim3$ GeV. When checking these data, we find possible enhancement structures existing in the corresponding invariant mass spectra with mass range from 2.4 GeV to 3 GeV.
For example, in the $e^+e^-\to\omega\pi^+\pi^-\pi^0$ process \cite{BaBar:2006vzy}, several enhancement structures existing in the energy range of 2.4 GeV are obvious. However, in the process of $e^+e^-\to\pi^+\pi^-\pi^0$, we cannot find conspicuous enhancement structure \cite{BESIII:2019gjz}. Thus, we still need more precise data.
In the following, we can find many narrow structures above 2.4 MeV in the process of $e^+e^-\to K_2^{*0}(1430)K^-\pi^+$ reported by the BaBar Collaboration \cite{BaBar:2011btv}.
The cross section of $e^+e^-\to\phi f_2'(1525)$ was also measured by the BaBar Collaboration. Here, there exists possible enhancement clusters around 2.7 GeV and a clear $J/\psi$ signal \cite{BaBar:2011btv}.
And then, the Belle Collaboration studied the process of $e^+e^-\to\phi\pi^+\pi^-$ and $e^+e^-\to\phi f_0(980)$ \cite{Belle:2008kuo}, where the event accumulation around 2.4 GeV and 2.6 GeV can be found.
We conjecture that these enhancement structures with low significance can be due to these unknown light unflavored vector mesons.

At present, our knowledge of higher states of the light unflavored vector meson family, which have the masses above 2.4 GeV, is obviously absent. This research status causes us to reinforce the investigation of these higher states of the light unflavored vector meson family, which becomes a central task of this work.

In this work, we apply the modified Godfrey-Isgur model \cite{Song:2015nia} to present the masses distribution of the discussed higher states of the light unflavored vector meson family, where their spatial wave functions can be gotten numerically, which can be as the input of the following calculation of their two-body Okubo-Zweig-lizuka (OZI) allowed decays. For performing the realistic estimate of these OZI allowed decays, the quark pair creation (QPC) model is adopted here \cite{Anisovich:2005wf,Roberts:1992js,Blundell:1996as}, which was applied to study the strong decay of different kinds of hadrons \cite{Yu:2011ta,Wang:2012wa,He:2013ttg,Ye:2012gu,Wang:2014sea,Chen:2015iqa,Guo:2019wpx,Pang:2017dlw,Pang:2018gcn,Wang:2020due}. We hope that our study on mass spectrum and decay width of the $\rho$, $\omega$, and $\phi$ mesons in $2.4\sim 3$ GeV can provide valuable information to the future experimental explorations.

As emphasized in Ref. \cite{BESIII:2020nme}, the BESIII experiment at the BEPCII is still an ideal platform to hunt for light hadrons. We have reason to believe that the present work can attract the BESIII interest of studying this topic in an experiment. In addition, the running of Belle II with the initial state radiation method will be a potential experiment to finding these discussed light unflavored vector mesons.

This paper is organized as follows. After the Introduction in Sec. \ref{sec1}, we present the spectroscopy behavior by the modified Godfrey-Isgur (MGI) model and the QPC model, where the mass spectrum and two-body OZI allowed strong decays of these discussed higher states of light unflavored vector meson are given in Sec. \ref{sec2}. Finally, the paper ends with a short summary in Sec. \ref{sec4}.

\section{Spectroscopy behavior}\label{sec2}

\begin{figure}[hptb]
\begin{center}
	\scalebox{0.9}{\includegraphics[width=\columnwidth]{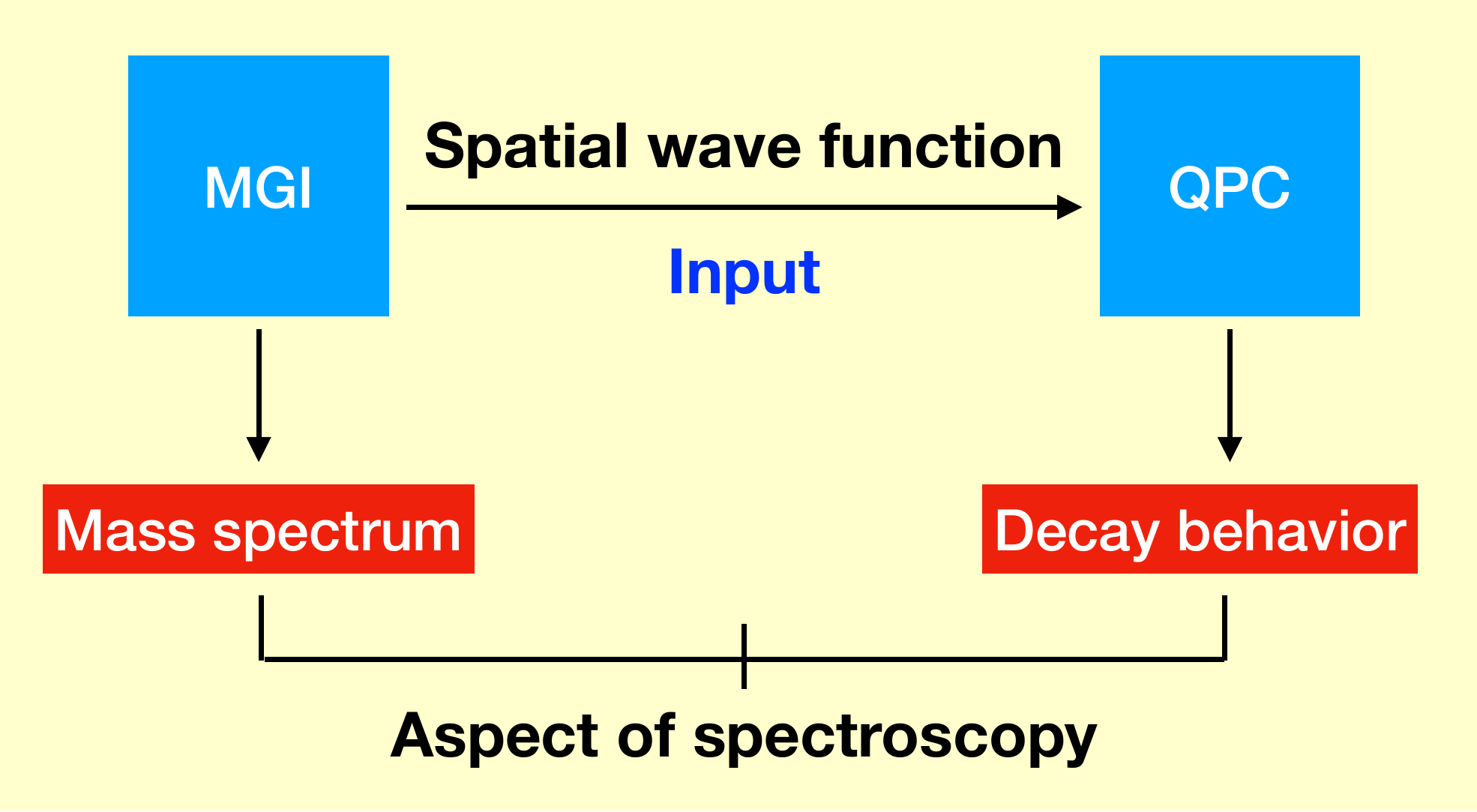}}
	\caption{The procedure of presenting the aspect of light unflavored vector  meson spectroscopy.} \label{procedure}
\end{center}
\end{figure}

There were different approaches to study the meson spectroscopy, which include lattice QCD~\cite{Dudek:2010wm,Dudek:2011tt}, Regge trajectories~\cite{Pang:2015eha,He:2013ttg,Badalian:2002xy}, nonrelativistic potential models~\cite{Eichten:1974af,Eichten:1978tg,Close:2005se}, QCD string approaches~\cite{Badalian:2006sg,Badalian:2008sv}, relativistic potential models~\cite{Ebert:2009ub,Ebert:2009ua}, and so on. In this work, we adopt the MGI model \cite{Song:2015nia} by replacing the linear potential with the screened potential in the Godfrey-Isgur (GI) model \cite{Godfrey:1985xj}. Here, the screened potential indicated by the lattice studies \cite{Laermann:1986pu,Born:1989iv} was employed in Refs. \cite{Chao:1992et,Ding:1993uy,Ding:1995he}, which is also inspired by the the coupled-channel effects \cite{Li:2009zu,Duan:2021alw}. A successful application of the MGI model is to depict the mass spectrum of these observed charmed-strange mesons \cite{Song:2015nia}, where the masses of the most observed $D_s$ mesons can be understood well. Later, the MGI model was further adopted to perform the mass spectrum analysis of charmed mesons \cite{Song:2015fha}, charmonia \cite{Wang:2019mhs,Wang:2020prx}, and bottomonia \cite{Wang:2018rjg}. Along this line, in this work we still apply the MGI model to discuss these higher states of the light unflavored vector meson family. Besides the mass spectrum, the spatial wave functions of these higher unflavored mesons could be extracted for getting the two-body OZI allowed strong decay widths via the QPC model. In Fig. \ref{procedure}, we show the procedure of how to obtain the whole aspect of these higher light unflavored vector meson spectroscopies.

\subsection{Mass spectrum}
The Hamiltonian depicting the interaction between quark and antiquark in the MGI model is
\begin{eqnarray}\label{hh}
\widetilde{H}=\sqrt{m_1^2+\mathrm{p^2}}+\sqrt{m_2^2+\mathrm{p^2}}+\widetilde{V}_{\mathrm{eff}}(\mathbf{p},\mathbf{r}),
\end{eqnarray}
where
$m_1$ and $m_2$ denote the mass of quark and antiquark, respectively. The effective potential of the $q\bar{q}$ interaction has the form \cite{Song:2015nia}
\begin{equation}\label{H}
\widetilde{V}_{\mathrm{eff}}(\mathbf{p},\mathbf{r})=\widetilde{H}^{\mathrm{OGE}}+\widetilde{H}^{\mathrm{conf}},
\end{equation}
where the first term $\widetilde{H}^{\mathrm{OGE}}$ is a $\gamma^{\mu}\otimes\gamma_{\mu}$ one-gluon-exchange potential, which is a short distance interaction, and the $\widetilde{H}^{\mathrm{conf}}$ in Eq.~(\ref{H}) is a long distance $1\otimes1$ confining interaction. The model includes smearing transformations and momentum-dependent terms. Firstly, the smearing transformations could be expressed as
\begin{equation}
\begin{split}
\widetilde{V}_{\rm eff}({\bf p},{\bf r})=&\int d^3{\bf r}^\prime \rho_{ij}({\bf r}-{\bf r}^\prime)V_{\rm eff}({\bf p},{\bf r}^\prime),\\
\end{split}
\end{equation}
where the $\rho_{ij}({\bf r}-{\bf r}^\prime)$ is the Gaussian smearing function with
\begin{equation}
\rho_{ij}({\bf r}-{\bf r}^\prime)=\frac{\sigma_{ij}^3}{\pi^{3/2}}e^{-\sigma_{ij}^2({\bf r}-{\bf r}^\prime)^2},
\end{equation}
\begin{equation}\label{sigmaij}
\sigma_{ij}^2=\sigma_0^2\left(\frac{1}{2}+\frac{1}{2}\left(\frac{4m_im_j}{(m_i+m_j)^2}\right)^4\right)+s^2\left(\frac{2m_im_j}{m_i+m_j}\right)^2.
\end{equation}
The parameters $\sigma_0$ and $s$ in Eq.~(\ref{sigmaij}) could be determined by the spin splits of the mesons.

The $\widetilde{H}^{\mathrm{OGE}}$ in Eq.~(\ref{H}) contains Coulomb, contact, tensor, and vector spin-orbit terms, i.e.,
\begin{equation}\label{HOGE}
\widetilde{H}^{\mathrm{OGE}}=\widetilde{G}_{12}(r)+\widetilde{V}^{\mathrm{cont}}(r)+\widetilde{V}^{\mathrm{tens}}(r)+\widetilde{V}^{\mathrm{sov}}(r).
\end{equation}
The Coulomb term $\widetilde{G}_{12}(r)$ is spin independent with definition
\begin{equation}
\begin{split}
\widetilde{G}_{ij}(r)=&\int d^3{\bf r}^\prime \rho_{ij}({\bf r}-{\bf r}^\prime)\left(\sum\limits_k-\frac{4\alpha_k }{3r}{\rm erf}(\gamma_kr)\right)\\
=&\sum\limits_k-\frac{4\alpha_k }{3r}{\rm erf}(\tau_kr),
\end{split}
\end{equation}
where $\alpha_k=(0.25,0.15,0.2)$ and $\gamma_k=(1/2,\sqrt{10}/2,\sqrt{1000}/2)$ with $k=1,2,3$ \cite{Godfrey:1985xj}, and the $\tau_k$ could be obtained by
\begin{equation}
\tau_k=\frac{1}{\sqrt{\frac{1}{\sigma_{ij}^2}+\frac{1}{\gamma_k^2}}}.
\end{equation}
Additionally, the semirelativistic correction with momentum-dependent factors is introduced. For the Coulomb term, the correction is
\begin{equation}
G_{12}(r)\to \left(1+\frac{p^2}{E\bar{E}}\right)^{1/2}G_{12}(r)\left(1+\frac{p^2}{E\bar{E}}\right)^{1/2}.
\end{equation}
For the spin-dependent terms, the semi-relativistic correction could be written as
\begin{equation}\label{Vialphabeta}
V^i_{\alpha\beta}\to \left(\frac{m_\alpha m_\beta}{E_\alpha E_\beta}\right)^{1/2+\epsilon_i}V_{\alpha\beta}\left(\frac{m_\alpha m_\beta}{E_\alpha E_\beta}\right)^{1/2+\epsilon_i}.
\end{equation}
The remaining terms in Eq.~(\ref{HOGE}) are spin-dependent potentials, which are defined as
\begin{equation}\label{Vcont}
\begin{split}
\widetilde{V}^{\mathrm{cont}}(r)=\frac{2{\bf S}_1\cdot{\bf S}_2}{3m_1m_2}\nabla^2\widetilde{G}_{12}^c(r),
\end{split}
\end{equation}
\begin{equation}\label{Vtens}
\begin{split}
\widetilde{V}^{\mathrm{tens}}(r)=&-\left(\frac{3{\bf S}_1\cdot{\bf r}{\bf S}_2\cdot{\bf r}/r^2-{\bf S}_1\cdot{\bf S}_2}{3m_1m_2}\right)\left(\frac{\partial^2}{\partial r^2}-\frac{1}{r}\frac{1}{\partial r}\right)\widetilde{G}_{12}^t(r),
\end{split}
\end{equation}
\begin{equation}\label{Vsov}
\begin{split}
\widetilde{V}^{\mathrm{sov}}(r)=&\frac{{\bf S}_1\cdot {\bf L}}{2m_1^2}\frac{1}{r}\frac{\partial\widetilde{G}_{11}^{\rm sov}(r)}{\partial r}+\frac{{\bf S}_2\cdot {\bf L}}{2m_2^2}\frac{1}{r}\frac{\partial\widetilde{G}_{22}^{\rm sov}(r)}{\partial r}\\
&+\frac{({\bf S}_1+{\bf S}_2)\cdot {\bf L}}{m_1m_2}\frac{1}{r}\frac{\partial\widetilde{G}_{12}^{\rm sov}(r)}{\partial r}.\\
\end{split}
\end{equation}
In Eqs.~(\ref{Vcont})-(\ref{Vsov}), the semirelativistic correction factors in Eq.~(\ref{Vialphabeta}) with $\epsilon_i=\epsilon_c$, $\epsilon_t$, and $\epsilon_{\rm sov}$ impacts the potentials $\widetilde{V}^{\mathrm{cont}}(r)$, $\widetilde{V}^{\mathrm{tens}}(r)$, and $\widetilde{V}^{\mathrm{sov}}(r)$, respectively\footnote{The spin-orbit and tensor interactions without smearing transformations may contain the $1/r^3$ term. But for $S$-wave states, the spin-orbit and tensor operators are zero, thus the spin-orbit and tensor potentials do not lead to singularities.}.

The second part of Eq.~(\ref{H}) contains a screened confinement potential and a scalar spin-orbit interaction, i.e.,
\begin{equation}\label{Hconf}
\widetilde{H}^{\mathrm{conf}}=\widetilde{S}_{12}+\widetilde{V}^{\mathrm{sos}}.
\end{equation}
In the MGI model, the screen effect \cite{Song:2015nia} is introduced, where we should make a replacement
\begin{equation}
br\to \frac{b(1-e^{-\mu r})}{\mu}.
\end{equation}
Then the $\widetilde{S}_{12}$ in Eq.~(\ref{Hconf}) is defined as
\begin{equation}\label{Sij}
\begin{split}
\widetilde{S}_{ij}(r)=&\int d^3{\bf r}^\prime \rho_{ij}({\bf r}-{\bf r}^\prime)\frac{b(1-e^{-\mu r})}{\mu}+c,
\end{split}
\end{equation}
where $b$ is the strength of the confinement, and the $\mu$ is a parameter to scale the screening effect. Then the scalar spin-orbit interaction could be obtained by
\begin{equation}
\begin{split}
\widetilde{V}^{\mathrm{sos}}(r)=&-\frac{{\bf S}_1\cdot {\bf L}}{2m_1^2}\frac{1}{r}\frac{\partial\widetilde{S}_{11}^{\rm sos}(r)}{\partial r}-\frac{{\bf S}_2\cdot {\bf L}}{2m_2^2}\frac{1}{r}\frac{\partial\widetilde{S}_{22}^{\rm sos}(r)}{\partial r},\\
\end{split}
\end{equation}
where the $\widetilde{S}_{11}^{\rm sos}(r)$ and $\widetilde{S}_{11}^{\rm sos}(r)$ contain semirelativistic correction factors in Eq.~(\ref{Vialphabeta}) with $\epsilon_i=\epsilon_{\rm sos}$.

In the above formula, $\mathbf{S_1}$ and $\mathbf{S_2}$ denote the spin of quark and antiquark, respectively. $\bm{L}$ indicates the orbital momentum between two quarks. More details of the MGI model can be found in Ref. \cite{Song:2015nia}.

\begin{table}[htbp]
\caption{The value of these parameters involved in the MGI model.}
\renewcommand\arraystretch{1.2}
\begin{tabular*}{86mm}{c@{\extracolsep{\fill}}cccc}
\toprule[1pt]\toprule[1pt]
Parameter                 & Value  & Parameter                 & Value  \\
\toprule[0.8pt]
$m_u$ (GeV)               & 0.22   & $m_d$ (GeV)               & 0.22   \\
$m_s$ (GeV)               & 0.424  & $b$ (GeV$^2$)             & 0.229  \\
$\epsilon_c$              & -0.164 & $\epsilon_{\mathrm{sos}}$ & 0.9728 \\
$\sigma_0$ (GeV)          & 1.8    & $s$ (GeV)                 & 3.88   \\
$\mu$ (GeV)               & 0.081  & $c$ (GeV)                 & -0.30 \\
$\epsilon_{\mathrm{sov}}$ & 0.262  & $\epsilon_t$              & 1.993  \\
\toprule[1pt]\toprule[1pt]
\end{tabular*}\label{para}
\end{table}

\begin{figure*}[hptb]
\centering
\includegraphics[width=15cm]{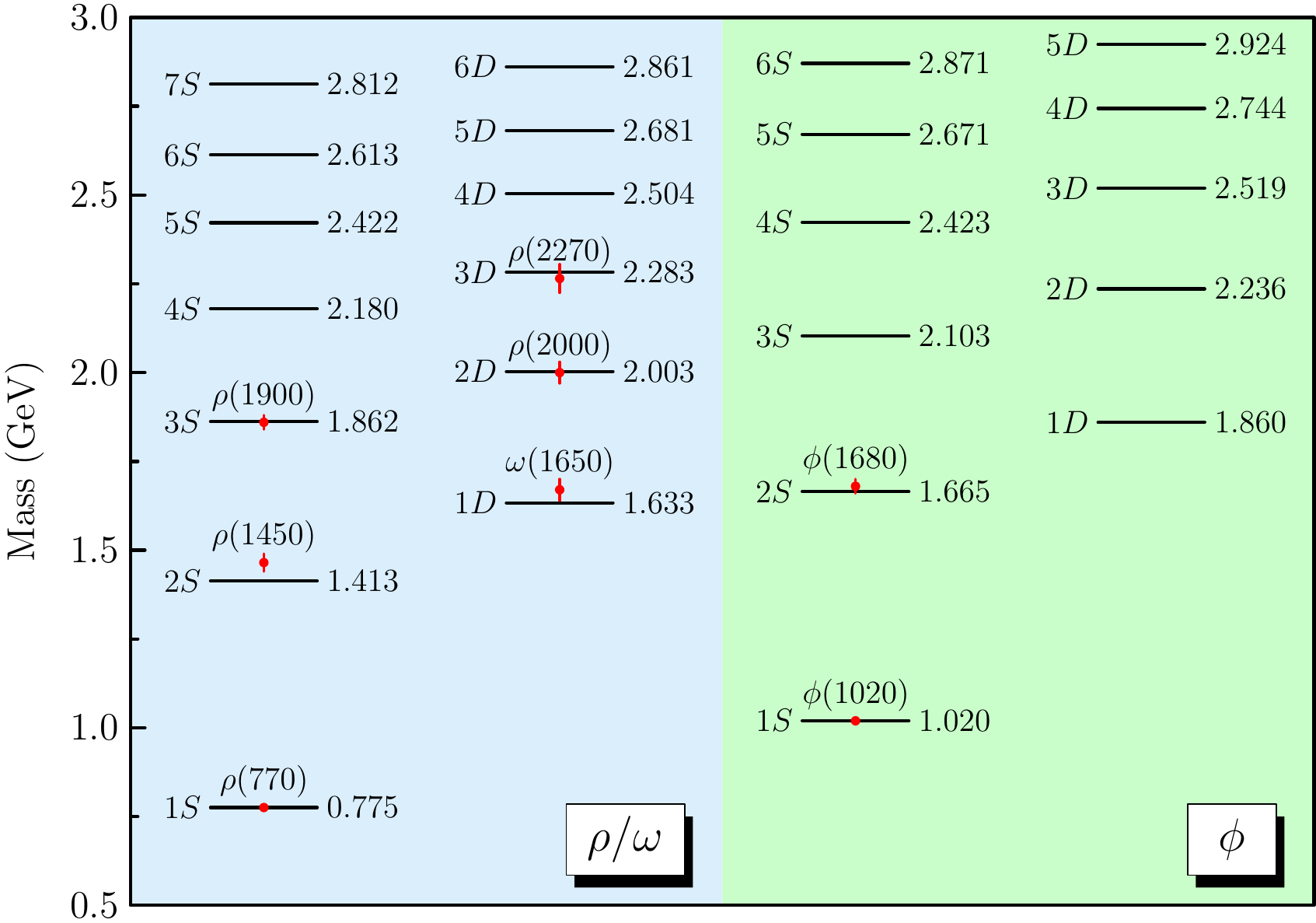}
\caption{The comparison of calculated results calculated in this work and the experimental data of mass of higher $\rho$, $\omega$ and $\phi$ states. Here, the red dots denote the experimental data from Refs. \cite{PDG:2020ssz,BaBar:2006vzy,Anisovich:2002su,Bugg:2004xu}.} \label{mass1}
\end{figure*}

The values of all parameters used in the MGI model are collected in Table. \ref{para}. With these preparations, we further present the mass spectrum of light unflavored vector mesons in Fig. \ref{mass1}. In Refs. \cite{Wang:2012wa,He:2013ttg,Wang:2020kte}, the Lanzhou group performed the study of mass spectrum and two-body OZI allowed decay of the mesons in the $\rho$, $\omega$, and $\phi$ family and discussed how to categorize these observed light vector $\rho$, $\omega$, and $\phi$ states below 2.4 GeV collected in Particle Data Group (PDG)  \cite{PDG:2020ssz} into the $\rho$, $\omega$, and $\phi$ family. Here, the MGI model was adopted, which was applied to study various light flavor systems, including kaon family \cite{Pang:2017dlw}, $\phi$ states below 2.6 GeV \cite{Pang:2019ttv}, and high spin light flavor mesons \cite{Pang:2018gcn}. When further exploring the property of these higher states of the $\rho$, $\omega$, and $\phi$ family, we still use the MGI model for presenting the mass spectrum analysis. Here, besides reproducing the masses of these low lying states, the masses of higher states are also obtained. Our calculated result explicitly shows that there exists accumulation of these $5S$, $6S$, and $7S$ states of $\rho$, $\omega$ and $\phi$ mesons in the mass range from 2.4 GeV to 3 GeV.

{In this work, we also calculate the dilepton decay widths of 
these discussed light unflavored vector mesons. The dilepton decay width of a unflavored meson can be expressed as~\cite{Godfrey:1985xj}
\begin{equation}\label{Gammaee}
\Gamma_{e^+e^-}=\frac{4\pi}{3}\alpha^2 m |{\cal M}|^2,
\end{equation}
where the $m$ is the mass of the vector meson. The ${\cal M}$ in Eq.~(\ref{Gammaee}) is the decay amplitude. For $S$-wave vector mesons, we employ ${\cal M}_\rho=\sqrt{6}V_{\rho}$, ${\cal M}_\omega=\sqrt{2/3}V_{\omega}$, and ${\cal M}_\phi=-\sqrt{4/3}V_{\phi}$ to express the decay amplitudes of the $\rho$, $\omega$, and $\phi$ states, respectively. In potential models, the factors $V_{\rho}$, $V_{\omega}$, and $V_{\phi}$ can be written as
\begin{equation}\label{Vrhoomegaphi}
V_{\rho/\omega/\phi}=m^{-2}\widetilde{m}^{1/2}(2\pi)^{-3/2}\int {\rm d}^3{\bf p}(4\pi)^{-1/2}\Phi(p)\left(\frac{m_1m_2}{E_1E_2}\right)^{1/2},
\end{equation}
where $m_1$ and $m_2$ are consistent quark masses in the meson, and $E_1$ and $E_2$ are energies of the quarks. The $\Phi(p)$ in Eq.~(\ref{Vrhoomegaphi}) is the radial part of the spatial wave function in the momentum space, which is extracted from the potential model. And the $\widetilde{m}$ is calculated by $\widetilde{m}=2\int{\rm d}^3{\bf p} E |\phi({\bf p})|^2$, where $\phi({\bf p})=\Phi(p)Y_{lm}(\Omega_{\bf p})$. Besides, we use $M_{\rho}^\prime=\sqrt{4/3}V_\rho^\prime$, $M_{\omega}^\prime=\sqrt{4/27}V_\omega^\prime$, and $M_{\phi}^\prime=-\sqrt{8/27}V_\phi^\prime$ to denote the decay amplitudes of the $D$-wave vector $\rho$, $\omega$, and $\phi$ mesons, respectively. The $V_\rho^\prime$, $V_\omega^\prime$, and $V_\phi^\prime$ could be expressed by
\begin{equation}
\begin{split}
V_{\rho/\omega/\phi}^\prime=&m^{-2}\widetilde{m}^{1/2}(2\pi)^{-3/2}\int {\rm d}^3{\bf p}(4\pi)^{-1/2}\phi(p)\\
&\times\left(\frac{m_1m_2}{E_1E_2}\right)^{1/2}\left(\frac{p}{E_1}\right)^2.
\end{split}
\end{equation}
With preparations above, the dilepton decay widths of these discussed light unflavored vector mesons are calculated and presented in Table \ref{Tab2}.
}

\begin{table}[htbp]
\centering
\caption{The dilepton decay widths of the discussed light unflavored vector mesons in units of keV.}
\label{Tab2}
\renewcommand\arraystretch{1.25}
\begin{tabular*}{86mm}{@{\extracolsep{\fill}}c@{\hskip\tabcolsep\vrule width 0.75pt\hskip\tabcolsep}cccccc}
\toprule[1.00pt]
\toprule[1.00pt]
States&$\rho(5S)$&$\rho(6S)$&$\rho(7S)$&$\rho(4D)$&$\rho(5D)$&$\rho(6D)$\\
\midrule[0.75pt]
$\Gamma_{e^+e^-}$&0.0377&0.0204&0.0195&0.0070&0.0052&0.0029\\
\midrule[0.75pt]
States&$\omega(5S)$&$\omega(6S)$&$\omega(7S)$&$\omega(4D)$&$\omega(5D)$&$\omega(6D)$\\
\midrule[0.75pt]
$\Gamma_{e^+e^-}$&0.0042&0.0023&0.0022&0.0008&0.0006&0.0003\\
\midrule[0.75pt]
States&$\phi(4S)$&$\phi(5S)$&$\phi(6S)$&$\phi(3D)$&$\phi(4D)$&$\phi(5D)$\\
\midrule[0.75pt]
$\Gamma_{e^+e^-}$&0.0492&0.0291&0.0162&0.0096&0.0058&0.0037\\
\bottomrule[1.00pt]
\bottomrule[1.00pt]
\end{tabular*}
\end{table}

In the following section, we pay more attention to their two-body OZI allowed strong decays, which can give important hints to the experimental search for them. Here, we need to emphasize that the numerical forms of the spatial wave function of these discussed light vector unflavored mesons can be obtained, which are as the input of the following discussion, by their decay behavior.

\subsection{Two-body OZI allowed strong decay}\label{sec3}

For calculating the two-body OZI allowed strong decay of these discussed higher unflavored vector meson states, we apply the QPC model, which is an effective approach to quantitatively estimate such physical quantity \cite{Anisovich:2005wf,Roberts:1992js,Blundell:1996as}.

In the QPC model, when a meson decay occurs, a quark-antiquark pair is created from the vacuum with the quantum number $J^{PC}=0^{++}$ and combines with the corresponding antiquark and quark in the initial meson to produce two final mesons. In Ref. \cite{Hayne:1981zy}, the authors introduced the wave functions of the mock states when calculating the transition amplitudes. In the QPC model, the wave function of a mock state is also adopted. In the decay process $A\to BC$, the wave function of the mock state $A$ is defined as \cite{Hayne:1981zy,Blundell:1996as}
\begin{equation}\label{mockstate}
\begin{split}
& \left|A\left(n_A^{2S_A+1}L_{AJ_AM_{J_A}}\right)\left(\bm{P}_A\right)\right\rangle\\
&\equiv \sqrt{2E_A}\sum\limits_{M_{L_A},M_{S_A}}
\left\langle L_AM_{L_A}S_AM_{S_A}|J_AM_{J_A}\right\rangle \\
&\quad\times\int d^3\bm{p}_A\psi_{n_AL_AM_{L_A}}(\bm{p}_A)\chi_{S_AM_{S_A}}^{12}\phi_A^{12}\omega_A^{12}\\&\quad\times
\left|q_1\left(\frac{m_1}{m_1+m_2}\bm{P}_A+\bm{p}_A\right)\bar{q}_2\left(\frac{m_2}{m_1+m_2}\bm{P}_A-\bm{p}_A\right)\right\rangle,
\end{split}
\end{equation}
where the $m_1$ and $m_2$ are masses of quark $q_1$ and antiquark $\bar{q_2}$, respectively. For the mock state wave function of the meson $B$ ($C$), we simply replace `$A$' by `$B$' (`$C$') in Eq.~(\ref{mockstate}). The $n_A$ in Eq.~(\ref{mockstate}) is the radial quantum number of the meson $A$. The $S_A$, $L_A$, and $J_A$ are spin, orbital angular momentum, and total angular momentum, respectively. Note that $\bm{p}_A=\frac{m_1\bm{p_1}-m_2\bm{p_2}}{m_1+m_2}$ is the relative momentum of the $q_1\bar{q}_2$ pair, where $\bm{p}_1$ and $\bm{p}_2$ are the momentum of $q_1$ and $\bar{q}_2$, respectively. Since the particles here are on shell, the integral variable ${\bm p}_A$ in Eq.~(\ref{mockstate}) is three dimensional. Note that $\bm{P}_A=\bm{p}_1+\bm{p}_2$ and $E_A$ are total momentum and total energy of the meson $A$, respectively. The $\psi_{n_AL_AM_{L_A}}(\bm{p}_A)$ is the spatial wave function in momentum space. The $\chi_{S_AM_{S_A}}^{12}$, $\phi_{A}^{12}$, and $\omega_A^{12}$ are the spin, flavor, and color wave functions, respectively.

The total decay width in the center-of-mass (CM) frame is given by
\begin{eqnarray}\label{Width}
\Gamma=\frac{\pi}{4}\frac{|\mathbf{P}|}{M_A^2}\sum\limits_{LS}\left|M^{LS}\right|^2.
\end{eqnarray}
Here, $\mathbf{P}$ is the momentum outgoing meson (meson $B$ or meson $C$). $M_A$ is the mass of meson $A$. $\bm{L}$ and $\bm{S}$ denote the relative orbital angular and total spin momentum between meson $B$ and $C$, respectively. $M^{LS}$ denotes the partial wave amplitude and is related to the helicity amplitude $M^{M_{J_A}M_{J_B}M_{J_C}}$ according to Jacob-Wick formula \cite{Jacob:1959at}. In the CM frame, the specific form of $M^{M_{J_A}M_{J_B}M_{J_C}}$ can be written as
\begin{eqnarray}\label{Amplitude}
\begin{split}
&M^{M_{J_A}M_{J_B}M_{J_C}}(\bm{P})\\&=\gamma\sqrt{8E_AE_BE_C}\sum\limits_{\mbox{$\fontsize{7pt}{7pt}{\begin{split}M_{L_A},M_{S_A},M_{L_B},\\M_{S_B},M_{L_C},M_{S_C},m\end{split}}$}}\langle L_AM_{L_A}S_AM_{S_A}|J_AM_{J_A}\rangle\;\\&\quad\times
\langle L_BM_{L_B}S_BM_{S_B}|J_BM_{J_B}\rangle\langle L_CM_{L_C}S_CM_{S_C}|J_CM_{J_C}\rangle\\
&\quad\times\langle 1m1-m|00\rangle\;\langle\chi_{S_BM_{S_B}}^{14}\chi_{S_CM_{S_C}}^{32}|\chi_{S_AM_{S_A}}^{12}\chi_{1-m}^{34}\rangle\\
&\quad\times\bigg[\langle \phi_B^{14}\phi_C^{32}|\phi_A^{12}\phi_0^{34}\rangle I(\bm{P},m_2,m_1,m_3)\\
&\quad+(-1)^{1+S_A+S_B+S_C}\langle \phi_B^{32}\phi_C^{14}|\phi_A^{12}\phi_0^{34}\rangle I(\bm{-P},m_2,m_1,m_3)\bigg],
\end{split}
\end{eqnarray}
where $\gamma$ as the strength of the quark-antiquark pair created from the vacuum is fixed to be 6.57 \cite{Pang:2019ttv}. $\phi_0$ denotes the flavor wave function of this quark-antiquark pair. Note that $I(\bm{P},m_2,m_1,m_3)$ is the momentum space integral
\begin{eqnarray}\label{Integration}
\begin{split}
I(\bm{P},m_2,m_1,m_3)=&\int d^3\bm{p}\psi_{n_BL_BM_{L_B}}^*\left(\frac{m_3}{m_1+m_3}\bm{P}+\bm{p}\right)\\
&\times \psi_{n_CL_CM_{L_C}}^*\left(\frac{m_3}{m_2+m_3}\bm{P}+\bm{p}\right)\\
&\times \psi_{n_AL_AM_{L_A}}(\bm{P}+\bm{p})\mathcal{Y}_1^m(\bm{p}),
\end{split}
\end{eqnarray}
where $\mathcal{Y}_1^m(\bm{p})$ is a solid harmonic that gives the momentum space distribution of the created quark-antiquark pair.

In our previous works, the QPC model was employed to deal with the two-body OZI allowed strong decays of light unflavored vector mesons, which include $\omega(2S)$, $\omega(1D)$, $\phi(2S)$, and $\phi(2D)$~\cite{Wang:2012wa} and $\rho(2S)$, $\rho(3S)$, $\rho(4S)$, $\rho(1D)$, $\rho(2D)$, and $\rho(3D)$~\cite{He:2013ttg}, where the experimental widths have been well reproduced.
 In addition, we predicted the partial and total widths of the $\phi(1D)$, $\omega(2D)$, $\phi(3S)$, and $\omega(3S)$ states~\cite{Wang:2012wa}. Until now, these excited light unflavored vector mesons below 2.4 GeV have been systemically studied. Thus, in the present work, we do not list the results of these low-lying states.

Focusing on the light unflavored vector mesons existing in the range of 2.4 $\sim$ 3 GeV, we continue to adopt the QPC model to estimate the strong decay widths, which is a main task of the following subsections.

\begin{figure*}[hptb]
\begin{center}
	\scalebox{1}{\includegraphics[width=17.2cm]{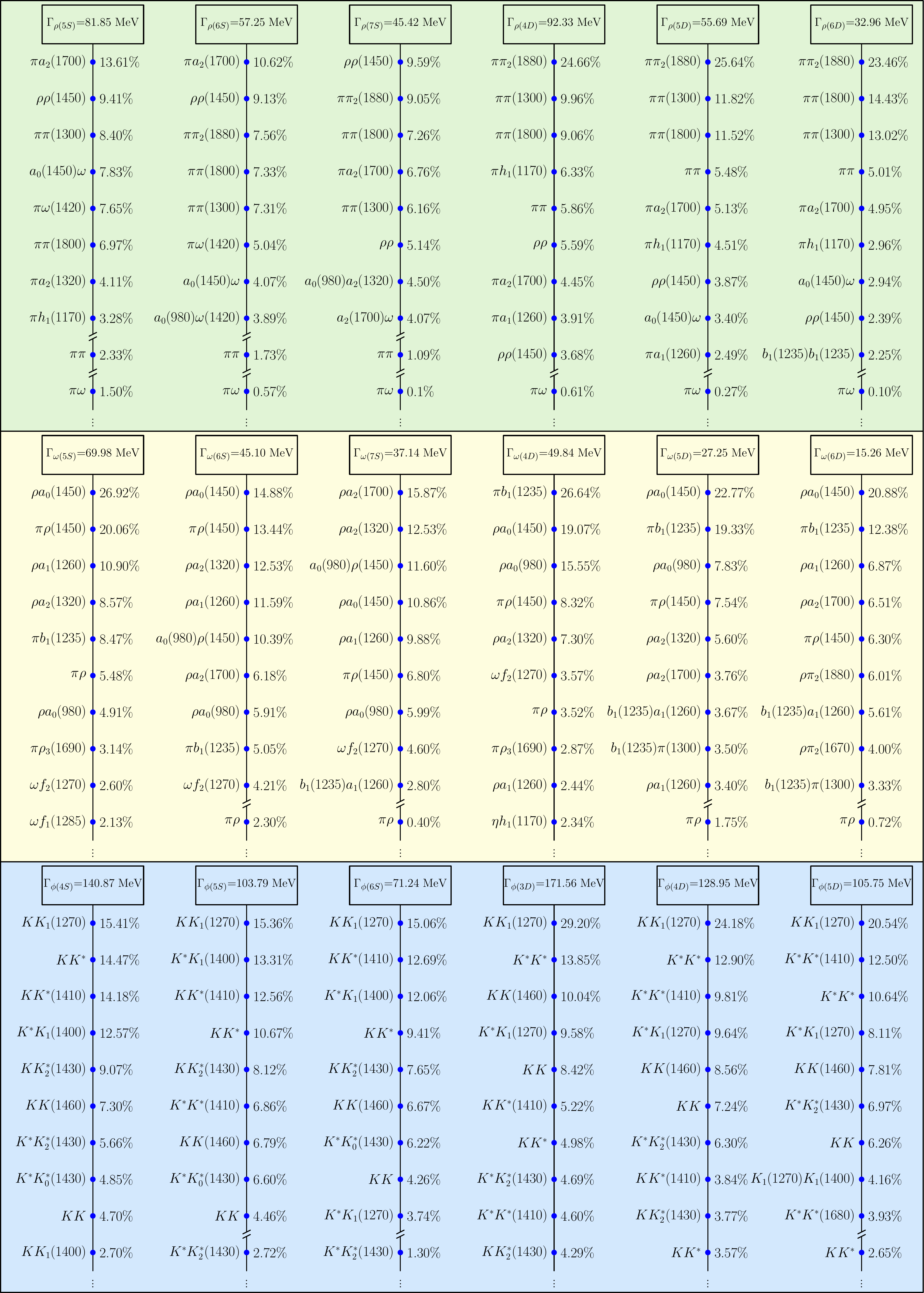}}
	\caption{The two-body strong decay behaviors of these discussed light vector mesons with the mass range from 2.4 GeV to 3 GeV. Here, we show their branching ratio of main channels and the total decay width.} \label{fig3}
\end{center}
\end{figure*}

\subsubsection{$\rho$ states}\label{sec3.1}

As shown in Fig. \ref{mass1}, there are six $\rho$ states [$\rho(5S)$, $\rho(6S)$, $\rho(7S)$, $\rho(4D)$, $\rho(5D)$, and $\rho(6D)$] existing in the mass range of  $2.4\sim 3$ GeV.

The obtained total and main partial decay widths of $\rho(5S)$ are shown in Fig. \ref{fig3}, which show that $\pi a_2(1700)$ is the dominant decay channel and the total width of $\rho(5S)$ can reach up to 81.85 MeV. In Fig. \ref{fig3}, we also list their several typical decay modes like $\pi\pi$ and $\pi\omega$ easily accessible at experiment. However, the branching ratios of these typical decays are not obvious. Of course, the $\rho(5S)$ state can also decay into an open-strange meson pair. However, our result shows that
their contribution to the total width is not significant. For example, $\rho(5S)\to KK_1(1270)$ and $K^*K_1(1270)$, have branching ratios of $0.43\%$ and $0.41\%$, respectively. Thus, in this work, we do not collect the decay channels relevant to the open-strange meson pair in Fig. \ref{fig3}.

The predicted decay behavior of $\rho(6S)$ is given in Fig. \ref{fig4}. Similar to the $\rho(5S)$ state, its dominant decay channel is also $\pi a_2(1700)$, while the $\rho\rho(1450)$, $\pi\pi_2(1880)$, $\pi\pi(1800)$ and $\pi\pi(1300)$ modes have sizable contributions to the total decay width, which is predicted to be around 57.25 MeV. Additionally, the particle widths of $KK_1(1270)$ and $K^*K_1(1270)$ are 0.18 MeV and 0.16 MeV, respectively, which are not listed in Fig. \ref{fig3}.

The calculation results for the $\rho(7S)$ are presented in  Fig. \ref{fig3}, where $\rho\rho(1450)$ and $\pi\pi_2(1880)$ are its dominant decay modes. In addition, the contribution of the $\pi\pi(1800)$, $\pi a_2(1700)$, $\pi\pi(1300)$ and $\rho\rho$ decay modes to total width cannot ignored. As the typical channels, $\pi\pi$ and $\pi\omega$ have the branching ratios of $1.09\%$ and $0.10\%$, respectively.

Accompanying with these three $S$-wave states, three corresponding $D$-wave states exist in the same mass range. In Fig. \ref{fig3}, the two-body OZI allowed strong decay behavior of $\rho(4D)$, $\rho(5D)$, and $\rho(6D)$ are given by presenting the total decay width and the branching ratios of their partial decays. The total decay widths of $\rho(4D)$, $\rho(5D)$, and $\rho(6D)$ are 92.33 MeV, 55.69 MeV, and 32.96 MeV, respectively, where their main decay modes are $\pi\pi_2(1880)$, $\pi\pi(1300)$ and $\pi\pi(1800)$. Since these three $D$-wave $\rho$ states have sizable $\pi\pi$ decay rates, the $e^+e^-\to \pi\pi$ process can be as the ideal channel to identify these $D$-wave $\rho$ states, which should stimulate experimentalists' interest in measuring the cross section of $e^+e^-\to \pi\pi$ with energy of collision up to 3 GeV.

\subsubsection{$\omega$ states}\label{sec3.2}

In the following, we check the two-body OZI allowed strong decay of six higher $\omega$ mesonic states, which include $\omega(5S)$, $\omega(6S)$, $\omega(7S)$, $\omega(4D)$, $\omega(5D)$, and $\omega(6D)$, where their total decay widths and partial decay widths estimated by the QPC model are shown in Fig. \ref{fig3}.

The total widths of $\omega(5S)$ and $\omega(6S)$, which
are contributed by the main decay channels like $\rho a_0(1450)$, $\pi\rho(1450)$, $\rho a_1(1260)$, and $\rho a_2(1320)$,
are 69.98 MeV and $45.10$ MeV, respectively.
It is worth noting that the typical mode $\pi\rho$ has sizable contribution to the total width. Thus, $e^+e^-$ annihilation into $\pi\rho$ is a perfect process to experimentally search for these $\omega$ states. Similar to the situation of the discussed $\rho$ states, $\omega(5S)$ and $\omega(6S)$  decaying into an open-strange meson pair are not significant.

The decay properties of $\omega(7S)$ we predicted are presented in Fig. \ref{fig3}, where its dominant decay mode is the $\rho a_2(1700)$ channel. Other main decay modes of $\omega(7S)$ include $\rho a_2(1320)$,  $a_0(980)\rho(1450)$, $\rho a_0(1450)$, and $\rho a_1(1260)$. These partial decay widths of $\omega(7S)$ into an open-strange meson pair are tiny in our calculation, which are not shown here.

Additionally, we also study three $D$-wave $\omega$ states existing in the mass range $2.4\sim 3$ GeV. The estimated total decay width and branching ratios of partial decay of $\omega(4D)$, $\omega(5D)$, and $\omega(6D)$ are listed in Fig. \ref{fig3}, where the obtained total width of $\omega(4D)$, $\omega(5D)$, and $\omega(6D)$ are 49.84 MeV, 27.25 MeV, and 15.26 MeV, respectively. These $D$-wave $\omega$ states mainly decay into $\pi b_1(1235)$, and $\rho a_0(1450)$. Other decay channels like $\pi\rho(1450)$ and $\rho a_0(980)$ also have a non-negligible effect on their total decay widths. Additionally, $\omega(4D)$ and $\omega(5D)$ all have sizable $\pi\rho$ decay rates, which are more significant than $\omega(6D)$.

\subsubsection{$\phi$ states}\label{sec3.3}

In this subsection, we discuss two-body strong decays of $\phi(4S)$, $\phi(5S)$, $\phi(6S)$, $\phi(3D)$, $\phi(4D)$, and $\phi(5D)$, which are also listed in Fig. \ref{fig3}.

From Fig. \ref{fig3}, we can notice that the main decay channels of $\phi(4S)$, $\phi(5S)$ and $\phi(6S)$ are $KK_1(1270)$, $KK^*$, $KK^*(1410)$, and $K^*K_1(1400)$, where the total decay widths of them are 140.87 MeV, 103.79 MeV, and 71.24 MeV, respectively.
Due to the sizable contribution of the channel $KK^*$ to the total decay widths of these discussed $\phi$ states, exploring $S$-wave $\phi$ mesonic states via the $e^+e^-\to KK^*$ process  is suggested.

In the same mass range, there are three $D$-wave $\phi$ mesons. In this work, we also estimate
their decay behaviors. $\phi(3D)$ has the total width of 171.56 MeV, where the branching ratio of its $KK_1(1270)$ decay channel is $29.20\%$, which is the main decay channel. Other modes including $K^*K^*$, $KK(1460)$, $K^*K_1(1270)$, and $KK$ are also significant. The predicted total width of $\phi(4D)$ is 128.95 MeV. The channels of $KK_1(1270)$, $K^*K^*$, $K^*K^*(1410)$, and $K^*K_1(1270)$ have the branching ratios of $24.18\%$, $12.90\%$, $9.81\%$, and $9.64\%$, respectively. And then, the obtained total decay width of $\phi(5D)$ is 95.12 MeV in this work. The results presented in Fig. \ref{fig3} further indicate that the branching ratio of the $\phi(5D)$ decays into $KK_1(1270)$, $K^*K^*(1410)$, $K^*K^*$, and $K^*K_1(1270)$ can reach up to $20.54\%$, $12.50\%$, $10.64\%$, and $8.11\%$, respectively.

Through the theoretical calculation given in this work, we  find  that the decay mode $K^*K_2^*(1430)$ of the six $\phi$ states discussed above has a significant contribution to their total decay widths. Thus, measuring the cross section of $e^+e^-\to K_2^*(1430)K\pi$ with energy of collision up to 3 GeV will be a promising approach to identify these highly excited $\phi$ states.
As shown in Fig. \ref{fig4} (c), indeed there are many obvious enhancement structures existing in the collision energy range of 2.4$\sim$2.7 GeV for $e^+e^-\to K_2^{*0}(1430)K^-\pi^+$, which may correspond to these discussed $\phi$ mesons. More precise data is expected in future experiments.

\section{Experimental evidence}\label{sec5}

\begin{figure}[hptb]
\begin{center}
	\scalebox{1}{\includegraphics[width=\columnwidth]{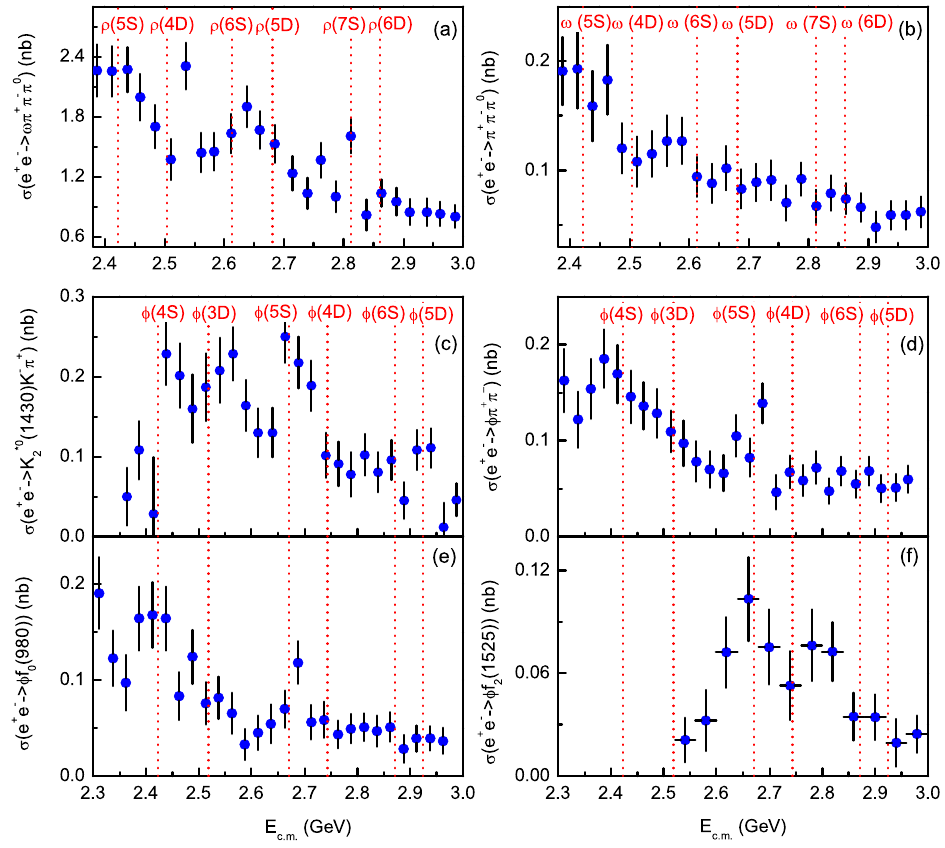}}
	\caption{A comparison of theoretical masses of these discussed light vector $\rho$, $\omega$ and $\phi$ mesons and the experimental data of $e^+e^-\to \omega\pi^+\pi^-\pi^0$ \cite{BaBar:2006vzy}, $e^+e^-\to \pi^+\pi^-\pi^0$ \cite{BESIII:2019gjz}, $e^+e^-\to K_2^{*0}(1430)K^-\pi^+$ \cite{BaBar:2011btv}, $e^+e^-\to\phi\pi^+\pi^-$ \cite{BaBar:2011btv}, $e^+e^-\to\phi f_0(980)$ \cite{BaBar:2011btv}, and $e^+e^-\to\phi f_2(1525)$ \cite{BaBar:2011btv}. These observed light vector $\rho$, $\omega$, and $\phi$ states below 2.4 GeV collected in PDG \cite{PDG:2020ssz} have been studied in our previous work~\cite{Wang:2012wa,He:2013ttg,Wang:2020kte}. Thus, here we do not list these light vector $\rho$, $\omega$, and $\phi$ states below 2.4 GeV.} \label{fig4}
\end{center}
\end{figure}

When having the theoretical results of masses of these higher states of light unflavored vector mesons, we can make comparison of our results with the experimental data as shown in Fig. \ref{fig4}. Firstly, focusing on the data of $e^+e^-\to \omega\pi^+\pi^-\pi^0$ \cite{BaBar:2006vzy}, which has a close relation with higher $\rho$ mesons due to the $G$-parity conservation, we find that the invariant mass distribution shows obvious enhancement structures like the event accumulation around 2.65 GeV. This broad structure overlaps with the predicted $\rho(6S)$ and $\rho(5D)$ states. There exists a jump point around 2.76 GeV, which may corresponds to $\rho(7S)$. And then, an unnotable fluctuation around 2.81 GeV just locates at the position of $\rho(6D)$. Additionally, a bump structure around 2.4 GeV should have relation with $\rho(5S)$, while around the $\rho(4D)$ state we can also find an obvious jump point.

The data of $e^+e^-\to \pi^+\pi^-\pi^0$ \cite{BESIII:2019gjz} is an ideal process to identify $\omega$ states. Although we mark these predicted $\omega$ states in the corresponding cross section data, we cannot identify the obvious bump structure corresponding to these states since the precision of the measured data is not enough. Thus, we have to wait for further precise measurements of the cross section of $e^+e^-\to \pi^+\pi^-\pi^0$.

In the following, we check four measured processes $e^+e^-\to K_2^{*0}(1430)K^-\pi^+$ \cite{BaBar:2011btv}, $e^+e^-\to\phi\pi^+\pi^-$ \cite{Belle:2008kuo}, $e^+e^-\to\phi f_0(980)$ \cite{Belle:2008kuo}, and $e^+e^-\to\phi f_2(1525)$ \cite{BaBar:2011btv}. For $e^+e^-\to K_2^{*0}(1430)K^-\pi^+$, there exists abundant event clusters. Since these predicted higher $\rho$, $\omega$ and $\phi$ states can decay into open-strange meson pairs, it is difficult to distinguish $\rho$, $\omega$ and $\phi$ states via the $e^+e^-\to K_2^{*0}(1430)K^-\pi^+$ channel. Further data analysis of $e^+e^-\to K_2^{*0}(1430)K^-\pi^+$ is an interesting research issue similar to the way adopted in Refs. \cite{Wang:2020kte,Wang:2021gle}.

Different from $e^+e^-\to K_2^{*0}(1430)K^-\pi^+$, $e^+e^-\to \phi f_0(980)$ can be the ideal process to identify these higher $\phi$ states. Two obvious structures around 2.4 GeV and 2.7 GeV may correspond to $\phi(4S)$ and $\phi(5S)$, respectively. Of course, we may also find an event accumulation around 2.5 GeV corresponding to $\phi(3D)$. For $e^+e^-\to\phi\pi^+\pi^-$, the structure at 2.41 GeV and the enhancement around 2.7 GeV may have relation to the $\phi(4S)$, $\phi(5S)/\phi(4D)$. For $e^+e^-\to\phi\pi^+\pi^-$, the broad structure around 2.65 GeV contains the information of the predicted $\phi(5S)/\phi(4D)$.

In fact, this simple comparison of the experimental data and masses of these discussed vector states only reflects some possible evidences of these predicted higher states of light unflavored vector meson. For further illustrating it,
in the following, we take two processes $e^+e^-\to \omega\pi^+\pi^-\pi^0$ and $e^+e^-\to \phi f_0(980)$ as examples to make further analysis.

In general, the cross section can be parametrized as the coherent sum of an $s$-dependent continuum amplitude and a resonant amplitude described by a Breit-Wigner function \cite{BESIII:2020kpr}, i.e.,
\begin{eqnarray}
\sigma(s)=\left |a\, e^{-b(\sqrt{s}-M_{th})}+\sum_{R}BW_{R}\left(\sqrt{s}\right)\,e^{i\phi_{R}}\right|^2,
\end{eqnarray}
where $a$ and $b$ are the continuum parameters, $M_{th}$ is the sum of the masses of the final state particles, and $\phi_{R}$ is the phase angle between the amplitudes.
The Breit-Wigner function is given by
\begin{eqnarray}
BW_{R}(\sqrt{s})=\frac{\sqrt{12\pi\Gamma^{e^+e^-}_{R}\mathcal{B}_{R}\Gamma^{\rm{tot}}_{R}}}{s-M_{R}^{2}+iM_{R}\Gamma^{\rm{tot}}_{R}},
\end{eqnarray}
where $M_R$, $\Gamma^{e^+e^-}_{R}$, and $\Gamma^{\rm{tot}}_R$ denote the mass, partial width to $e^+e^-$, and total width of the assumed intermediate resonance $R$. $\mathcal{B}_R$ is the
branching ratio for $R\to \rm{(final}$ $\rm{states)}$.

%%%%%%%%%%%%%%%%%%%%%%%%%%%%%%%%%%%%%%%%%%%%%%%%%%%%%%%%%%%%%%%%%
\begin{table}[htb]
  \centering
  \caption{The fitting parameters of depicting the Born cross section of $e^+e^-\to\omega\pi^+\pi^-\pi^0$.}\label{p1}
  \begin{tabular}{ccc}
  \toprule[1pt]
  \midrule[1pt]
  Parameters & $\quad$ & Solution \\
  \midrule[1pt]
  $\Gamma_{e^+e^-}\mathcal{B}(\rho(5S)\to\omega\pi^+\pi^-\pi^0)$ (eV) & $\quad$ & $3.38\pm0.96$\\
  $\Gamma_{e^+e^-}\mathcal{B}(\rho(6S)\to\omega\pi^+\pi^-\pi^0)$ (eV) & $\quad$ & $0.62\pm0.44$\\
  $\Gamma_{e^+e^-}\mathcal{B}(\rho(5D)\to\omega\pi^+\pi^-\pi^0)$ (eV) & $\quad$ & $0.86\pm0.52$\\
  $\Gamma_{e^+e^-}\mathcal{B}(\rho(6D)\to\omega\pi^+\pi^-\pi^0)$ (eV) & $\quad$ & $0.96\pm0.52$\\
  $\phi_{\rho(5S)}$ (rad)& $\quad$ & $4.76\pm0.21$\\
  $\phi_{\rho(6S)}$ (rad)& $\quad$ & $3.92\pm0.44$\\
  $\phi_{\rho(5D)}$ (rad)& $\quad$ & $5.52\pm0.35$\\
  $\phi_{\rho(6D)}$ (rad)& $\quad$ & $5.20\pm0.38$\\
  $a$ ($\times10^{-3}$ $\rm{GeV}^{-2}$) & $\quad$ & $-4.08\pm0.05$\\
  $b$ & $\quad$ & $0.54\pm0.01$\\
  $\chi^2/\rm{d.o.f}$ & $\quad$ & $0.60$\\
\midrule[1pt]
\bottomrule[1pt]
\end{tabular}
\end{table}
%%%%%%%%%%%%%%%%%%%%%%%%%%%%%%%%%%%%%%%%%%%%%%%%%%%%%%%%%%%%%%%%%%%%%%

%%%%%%%%%%%%%%%%%%%%%%%%%%%%%%%%%%%%%%%%%%%%%%%%%%%%%%%%%%%%%%%%%
\begin{table}[htb]
  \centering
  \caption{The fitting parameters of depicting the Born cross section of $e^+e^-\to\phi f_0(980)$.}\label{p2}
  \begin{tabular}{ccc}
  \toprule[1pt]
  \midrule[1pt]
  Parameters & $\quad$ & Solution \\
  \midrule[1pt]
  $\Gamma_{e^+e^-}\mathcal{B}(\phi(5S)\to \phi f_0(980))$ (eV) & $\quad$ & $1.36\pm0.0.31$\\
  $\Gamma_{e^+e^-}\mathcal{B}(\phi(6S)\to\phi f_0(980))$ (eV) & $\quad$ & $0.94\pm0.32$\\
  $\phi_{\phi(5S)}$ (rad)& $\quad$ & $4.32\pm0.21$\\
  $\phi_{\phi(6S)}$ (rad)& $\quad$ & $3.86\pm0.17$\\
  $a$ ($\times10^{-3}$ $\rm{GeV}^{-2}$) & $\quad$ & $-0.92\pm0.004$\\
  $b$ & $\quad$ & $1.92\pm0.09$\\
  $\chi^2/\rm{d.o.f}$ & $\quad$ & $0.49$\\
\midrule[1pt]
\bottomrule[1pt]
\end{tabular}
\end{table}
%%%%%%%%%%%%%%%

\begin{figure}[hptb]
\begin{center}
	\scalebox{1}{\includegraphics[width=\columnwidth]{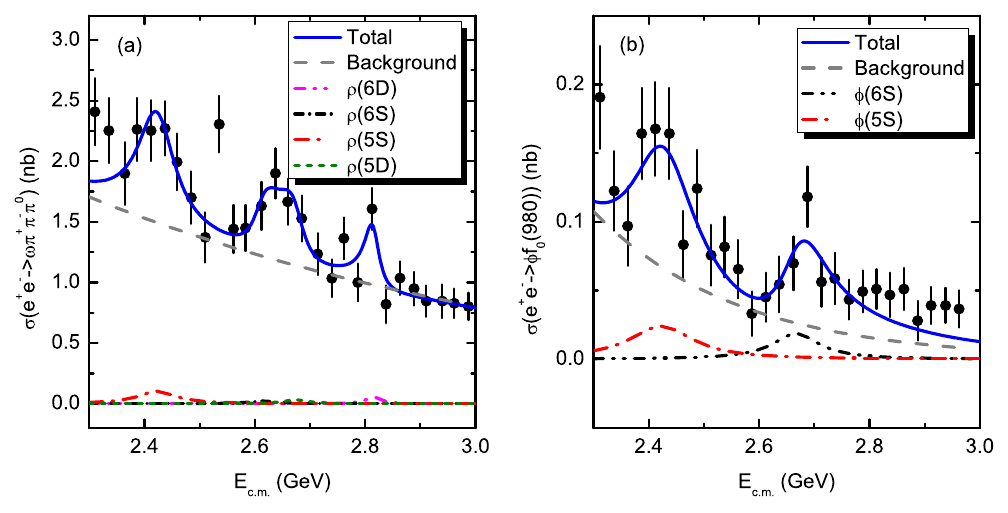}}
	\caption{The fitted result of the cross section of (a) $e^+e^-\to\omega\pi^+\pi^-\pi^0$ and (b) $e^+e^-\to\phi f_0(980)$. The black dots with the error bar are experimental data \cite{BaBar:2006vzy,BaBar:2011btv}.} \label{fig5}
\end{center}
\end{figure}

With this preparation, we first fit the measured cross section of $e^+e^-\to \omega\pi^+\pi^-\pi^0$ \cite{BaBar:2006vzy}. The comparison in Fig. \ref{fig4} (a) indicates that
$\rho(5S)$, $\rho(6S)$, $\rho(5D)$ and $\rho(6D)$ should be considered in our fit, where the resonance parameters for these vector states are taken from our calculation given in Sec. \ref{sec2} and the fitted parameters are listed in Table \ref{p1}.
As shown in Fig. \ref{fig5} (a), we can depict the data of the cross section of $e^+e^-\to \omega\pi^+\pi^-\pi^0$ \cite{BaBar:2006vzy}, by which the values of $\Gamma_R^{e^+e^-}\mathcal{B}_R$ for these considered light unflavored vector mesons can be obtained.
%The spectroscopy study in Sec. \ref{sec2} makes us to calculate partial width of these discussed light unflavored vector mesons decaying into $e^+e^-$, which are collected in Table \ref{Tab2}. Thus, combining with the fitting results and the calculated partial widths to $e^+e^-$, we further extract the branching ratios of $\rho(5S)$, $\rho(6S)$, $\rho(5D)$ and $\rho(6D)$ into $ \omega\pi^+\pi^-\pi^0$, which are 8.97\%, 3.04\%, 16.54\% and 33.10\%, respectively.
This information is valuable to further study around these light unflavored vector mesons.

Along this line, we may study the $e^+e^-\to \phi f_0(980)$ process \cite{BaBar:2011btv}. The fitted result can be found in Fig. \ref{fig5} (b). When considering the contributions from $\phi(5S)$ and $\phi(6S)$, we can describe the experimental data well, where the fitting parameters are listed in Table \ref{p2}. %Similar to the treatment of $e^+e^-\to \omega\pi^+\pi^-\pi^0$, we may extract the branching ratios of the $\phi(5S)$ and $\phi(6S)$ decays into $\phi f_0(980)$, which are 2.76\% and 3.23\% respectively.

We expect more precise data, which can be applied to identify these discussed higher light unflavored vector mesons.

\section{summary}\label{sec4}

In the past decades, the process of $e^+e^-$ annihilation into light mesons can be as an  ideal platform to probe the light vector mesonic state. A typical example is the observation of $Y(2175)$, which had stimulated theorists' interest in revealing its inner structure \cite{Wang:2021gle,Wang:2006ri,Ding:2007pc,MartinezTorres:2008gy,Chen:2008ej,Klempt:2007cp}. In recent years, BESIII, as the main force of studying light hadron spectroscopy, has accumulated more and more data of $e^+e^-$ annihilation into light mesons \cite{BESIII:2019gjz,BESIII:2021bjn,BESIII:2021yam,BESIII:2018ldc,BESIII:2019ebn,BESIII:2020gnc,BESIII:2020kpr,BESIII:2020vtu,BESIII:2020xmw}. Especially, these experimental measurements from BESIII make that constructing light unflavored vector mesons around 2 GeV become possible \cite{BESIII:2018ldc,BESIII:2020vtu,Wang:2020kte,Wang:2021gle,BESIII:2020xmw}.

Although big progress has been made by both experimentalists and theorists, our knowledge of these light unflavored vector mesons existing in the mass range of $2.4\sim 3$ GeV is not abundant, which inspires our attention.

In this work, we obtain the masses and decay widths of these discussed higher $\rho$, $\omega$, and $\phi$ states in the $2.4\sim 3$ GeV mass range and
make a rough comparison with the present experimental data as shown in Fig. \ref{fig4}. Although the present precision of experimental data is not enough to conclude whether these event clusters can be identified as these predicted states, we still want to show this possible evidence. With the improvement of experimental precision, we suggest our experimental colleagues  pay attention to these predicted higher vector states.

As emphasized in the recent white paper released by BESIII \cite{BESIII:2020nme}, the study around light hadrons will still be one of the tasks of the BESIII experiment in the next ten years. According to our study, we strongly suggest our experimental colleagues  pay attention to the 
$2.4\sim 3$ GeV energy range of collision, which has a close relation to these predicted light unflavored vector meson states.

\section*{Acknowledgments}
L.-M.W. would like to thank Lanzhou Center for Theoretical Physics to support his stay at Lanzhou University, where this work was finished. Additionally, L.-M. W. would like to thank Qin-Song Zhou for his help of fitting the experimental data.
This work is supported by the China National Funds for Distinguished Young Scientists under Grant No. 11825503, National Key Research and Development Program of China under Contract No. 2020YFA0406400, the 111 Project under Grant No. B20063, the National Natural Science Foundation of China under Grant No. 12047501, and
the projects funded by Science and Technology Department of Qinghai Province Project No. 2020-ZJ-728.

\end{document}